\documentclass[twocolumn]{aastex631}
\usepackage{graphicx}
\usepackage{tablefootnote}
\usepackage[T1]{fontenc} 
\usepackage[utf8]{inputenc} 
\graphicspath{{images/}}
\usepackage{amsmath}

\newcommand{\Msun}{\mbox{$\text{M}_{\sun}$}}
\newcommand{\Rsun}{\mbox{$\text{R}_{\sun}$}}
\newcommand{\Mjup}{\mbox{$\text{M}_{\rm Jup}$}}
\newcommand{\Rjup}{\mbox{$\text{R}_{\rm Jup}$}}
\newcommand{\Mearth}{\mbox{$\text{M}_{\oplus}$}}
\newcommand{\Rearth}{\mbox{$\text{R}_{\oplus}$}}
\newcommand{\PSUAA}{Department of Astronomy \& Astrophysics, 525 Davey Laboratory, The Pennsylvania State University, University Park, PA, 16802, USA}
\newcommand{\PSUCEHW}{Center for Exoplanets and Habitable Worlds, 525 Davey Laboratory, The Pennsylvania State University, University Park, PA, 16802, USA}

\newcommand{\UA}{Steward Observatory, University of Arizona, 933 N.\ Cherry Ave, Tucson, AZ 85721, USA}

\newcommand{\JPL}{Jet Propulsion Laboratory, California Institute of Technology, 4800 Oak Grove Drive, Pasadena, California 91109}

\newcommand{\UCI}{Department of Physics \& Astronomy, The University of California, Irvine, Irvine, CA 92697, USA}
\newcommand{\Carleton}{Carleton College, One North College St., Northfield, MN 55057, USA}

\newcommand{\UT}{Department of Astronomy and McDonald Observatory, University of Texas at Austin, 2515 Speedway, Austin, TX 78712, USA}

\begin{document}

\title{TOI-5344~b: A Saturn-like planet orbiting a super-Solar metallicity M0 dwarf}

\author[0000-0002-7127-7643]{Te Han} 
\affiliation{\UCI}

\author[0000-0003-0149-9678]{Paul Robertson}
\affil{\UCI}

\author[0000-0001-8401-4300]{Shubham Kanodia}
\affiliation{Earth and Planets Laboratory, Carnegie Institution for Science, 5241 Broad Branch Road, NW, Washington, DC 20015, USA}

\author[0000-0003-4835-0619]{Caleb Ca\~nas}
\affil{NASA Postdoctoral Fellow}
\affil{NASA Goddard Space Flight Center, 8800 Greenbelt Road, Greenbelt, MD 20771, USA}

\author[0000-0002-9082-6337]{Andrea S.J.\ Lin}
\affil{\PSUAA}
\affil{\PSUCEHW}

\author[0000-0001-7409-5688]{Guðmundur Stefánsson} 
\affil{NASA Sagan Fellow}
\affil{Department of Astrophysical Sciences, Princeton University, 4 Ivy Lane, Princeton, NJ 08540, USA}

\author[0000-0002-2990-7613]{Jessica E. Libby-Roberts}
\affil{\PSUAA}
\affil{\PSUCEHW}

\author[0000-0002-2401-8411]{Alexander Larsen}
\affil{Department of Physics \& Astronomy, University of Wyoming, Laramie, WY 82070, USA}

\author[0000-0002-4475-4176]{Henry A.~Kobulnicky}
\affil{Department of Physics \& Astronomy, University of Wyoming, Laramie, WY 82070, USA}

\author[0000-0001-9596-7983]{Suvrath Mahadevan}
\affil{\PSUAA}
\affil{\PSUCEHW}


\author[0000-0003-4384-7220]{Chad F.\ Bender}
\affil{\UA}

\author[0000-0001-9662-3496]{William~D.~Cochran}
\affil{McDonald Observatory and Center for Planetary Systems Habitability, The University of Texas, Austin, TX 78712, USA}

\author[0000-0002-7714-6310]{Michael Endl}
\affil{\UT}

\author[0000-0002-0885-7215]{Mark E. Everett}
\affil{NSF’s National Optical-Infrared Astronomy Research Laboratory, 950 N. Cherry Avenue, Tucson, AZ 85719, USA}

\author[0000-0002-5463-9980]{Arvind F.\ Gupta}
\affil{\PSUAA}
\affil{\PSUCEHW}

\author[0000-0003-1312-9391]{Samuel Halverson}
\affil{\JPL}

\author[0000-0002-1664-3102]{Fred Hearty}
\affil{\PSUAA}
\affil{\PSUCEHW}

\author[0000-0002-0048-2586]{Andrew Monson}
\affil{\UA}

\author[0000-0001-8720-5612]{Joe P.\ Ninan}
\affil{Department of Astronomy and Astrophysics, Tata Institute of Fundamental Research, Homi Bhabha Road, Colaba, Mumbai 400005, India}


\author[0000-0001-8127-5775]{Arpita Roy}
\affil{Space Telescope Science Institute, 3700 San Martin Drive, Baltimore, MD 21218, USA}
\affil{Department of Physics and Astronomy, Johns Hopkins University, 3400 N Charles St, Baltimore, MD 21218, USA}

\author[0000-0002-4046-987X]{Christian Schwab}
\affil{School of Mathematical and Physical Sciences, Macquarie University, Balaclava Road, North Ryde, NSW 2109, Australia}

\author[0000-0002-4788-8858]{Ryan C. Terrien}
\affil{\Carleton}

\begin{abstract}
We confirm the planetary nature of TOI-5344~b as a transiting giant exoplanet around an M0 dwarf star. TOI-5344~b was discovered with the Transiting Exoplanet Survey Satellite photometry and confirmed with ground-based photometry (the Red Buttes Observatory 0.6m telescope), radial velocity (the Habitable-zone Planet Finder), and speckle imaging (the NN-Explore Exoplanet Stellar Speckle Imager). TOI-5344~b is a Saturn-like giant planet ($\rho$ = 0.80$^{+0.17}_{-0.15}$ g cm$^{-3}$) with a planetary radius of 9.7 $\pm \ 0.5 \ \Rearth$ (0.87 $\pm \ 0.04 \ \Rjup$) and a planetary mass of 135$^{+17}_{-18} \ \Mearth$ (0.42$^{+0.05}_{-0.06} \ \Mjup$). It has an orbital period of $3.792622 \pm 0.000010$ days and an orbital eccentricity of 0.06$^{+0.07}_{-0.04}$. We measure a high metallicity for TOI-5344 of [Fe/H] = $0.48 \pm 0.12$, where the high metallicity is consistent with expectations from formation through core accretion. We compare the metallicity of the M-dwarf hosts of giant exoplanets to that of M-dwarf hosts of non-giants ($\lesssim 8\ \Rearth$). While the two populations appear to show different metallicity distributions, quantitative tests are prohibited by various sample caveats.
\end{abstract}



\section{Introduction}
Giant Exoplanets around M-dwarf Stars (GEMS) are planets with $ 8 \ \Rearth \lesssim R_p \lesssim 15 \ \Rearth$ and $M_p \sin i > 80 \ \Mearth$. While the formation mechanism of GEMS remains unclear, the popular core accretion theory \citep{Laughlin04, Ida05} predicts a low occurrence rate of GEMS. The small M-dwarf host masses generally correlate with low-mass protoplanetary disks, restricting the accretion of enough material to form giant planets. The gravitational instability mechanism \citep[GI;][]{Boss06} describes another scenario in which giant planets form by gravitational collapsing at large separation to form GEMS. This requires a massive disk-to-star ratio ($\simeq$ 10 \%) and also favors a low occurrence rate of GEMS. In fact, GEMS are so rare that previous radial velocity (RV) studies \citep{Endl06, Kovecs13, Sabotta21, Pinamonti22} were only able to set an upper limit of $\sim$1-3$\%$ occurrence rate for them\footnote{The definition of giant planets orbiting M-dwarf stars may slightly vary in different studies.}.  The Transiting Exoplanet Survey Satellite \citep[TESS;][]{Ricker15} has finally enabled a nearly all-sky search for transiting planets, of which dozens of new transiting GEMS have been discovered. \cite{Gan23} conclude a hot Jupiter ($7 \ \Rearth \leq R_p \leq 2 \ \Rjup$, $0.8 \leq P_b \leq 10$ days) occurrence rate of 0.27 $\pm$ 0.09\% around early-type M-dwarfs ($0.45 \ \Msun \leq M_* \leq 0.65 \ \Msun$) based on a box least squares search over 60,819 M dwarf TESS light curves. An even lower occurrence rate of 0.194 $\pm$ 0.072\% is reported by \cite{Bryant23} for the entire M dwarf range.

Mass measurements of the transiting objects are essential in confirming their nature as planets. In particular, transiting GEMS allow for precise mass measurement from RV observations, allowing a thorough study of the relations between the planetary mass and other stellar or planetary parameters. In this work, we confirm the planetary nature of TOI-5344~b, a Saturn-like planet orbiting a metal-rich M-dwarf star. With the addition of TOI-5344~b, the number of transiting GEMS is approaching 20, enabling statistics on this small population. We analyze the known transiting GEMS in various parameter spaces and highlight the planet mass-stellar metallicity relation. GEMS' hosts have been noted to have super-Solar metallicity by several studies \citep{Gan22, Kanodia22, Kagetani23}. This tendency is consistent with the mass budget argument \citep{Pollack96}, which requires high dust content in the protoplanetary disk (a result of high stellar metallicity) to form $\sim 10 \ \Mearth$ cores fast enough to initiate runaway gas accretion. We show the M-dwarf hosts of giant planets appear to have different metallicity distribution than the M-dwarf hosts of non-giants, which we define for this work to refer to planets that have radii $\lesssim 8\ \Rearth$. 



TOI-5344~b was discovered using TESS and confirmed with ground-based photometry, high contrast speckle imaging, and precise RVs. In Section~\ref{sec:observations}, we detail these observations. Stellar parameters are derived in Section~\ref{sec:stellar}, and a joint fit of photometry and RVs is presented in Section~\ref{sec:joint_fit}. We discuss TOI 5344~b in the GEMS parameter space in Section~\ref{sec:discussion} and conclude our findings in Section~\ref{sec:summary}. We also quantitatively compare our results to that of \cite{hartman23} on the same planet in each relevant section. 



\section{Observations} \label{sec:observations}
\subsection{TESS Photometry} \label{sec:tess}
TESS observed TOI-5344 (TIC 16005254; Gaia DR3 52359538285081728) in Sectors 43 and 44 from 2021 September 16 to 2021 November 6 at 600s cadence in the full-frame images (FFIs) and detected a total of 10 transits. The planet candidate TOI-5344~b was identified by the TESS Faint Star Search \cite{faint-star} using data products from the Quick Look Pipeline \cite[QLP;][]{QLP, Kunimoto22}. 

\begin{figure*}
    \centering
    \includegraphics[width=\linewidth]{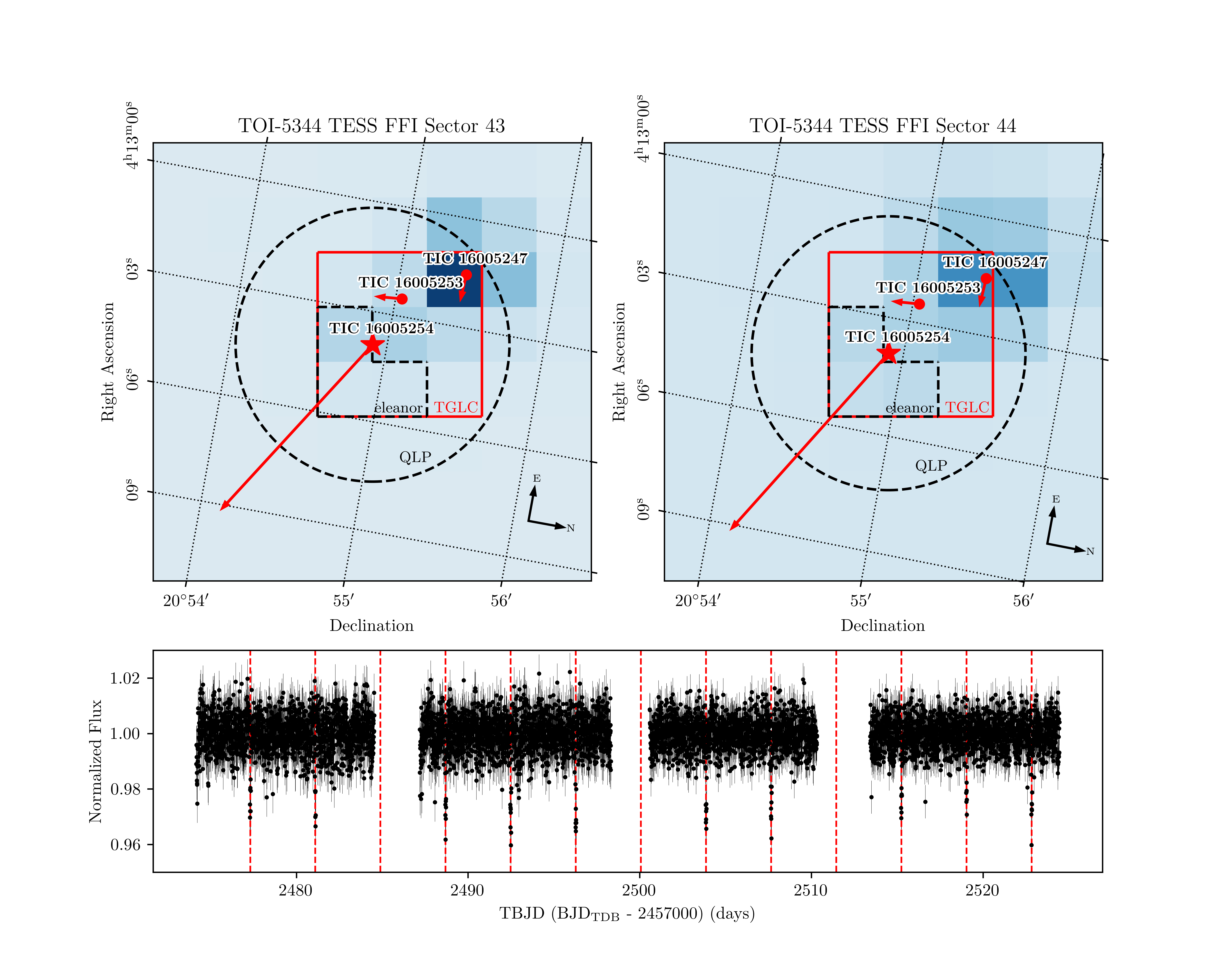}
    \caption{TESS photometry of TOI-5344. Top: TESS Full-Frame Images (FFIs) for Sector 43 and Sector 44 where Gaia stars (red) and their propagated positions after 2000 years (arrows) are plotted. The aperture used by TGLC is shown in red squares; the apertures used by \texttt{eleanor} and QLP are shown in black dashed L shapes and circles, respectively. Note that the TGLC aperture is applied to the decontaminated image instead of the shown raw FFI. Bottom: TGLC calibrated aperture light curve of TOI-5344 for Sector 43 and Sector 44. A total of ten observed transits are marked with red dashed lines. }
    \label{fig:tess}
\end{figure*}

We compared TESS FFI light curves from three pipelines: \texttt{eleanor} \citep{eleanor}, the Quick-Look Pipeline \citep[QLP;][]{QLP}, and TESS-Gaia Light Curve \citep[TGLC;][]{TGLC}. Each TESS FFI light curve was jointly fit with ground-based photometry and radial velocity with a free-floating TESS dilution factor. Only the dilution factor of TGLC calibrated aperture light curve (\texttt{cal\_aper\_flux}) fit is $\sim 1 \sigma$ from 1 ($D_{\text{TGLC}}$ = 1.075$^{+0.068}_{-0.065}$). In comparison, \texttt{CORR\_FLUX} of \texttt{eleanor} results in $D_{\texttt{eleanor}}$ = 0.88 $\pm$ 0.05; \texttt{KSPSAP\_FLUX} of QLP results in $D_{\text{QLP}}$ = 0.88 $\pm$ 0.06. Figure~\ref{fig:tess} shows the aperture used by all three light curves. Unlike \texttt{eleanor} and QLP which apply the shown apertures directly on the FFI, the TGLC aperture is applied to the decontaminated images where all nearby stars' point spread functions (PSFs) are modeled and removed. Therefore, the aperture flux of TGLC has much less contamination. To fit all stars' PSFs, the residual image of TGLC is minimized to zero, so there are similar chances of overestimation ($D < 1$) and underestimation ($D > 1$) of a single star's total flux. This explains why $D_{\text{TGLC}}$ may exceed 1. However, methods without contamination removal usually overestimate a star's total flux, so they can only have $D \leq 1$. The contamination removal allows the TGLC joint fit to imply a $\sim 7\%$ higher precision on the planet radius to stellar radius ratio than either \texttt{eleanor} or QLP joint fit results. We adopted the TGLC aperture light curve for our analysis because it removes contamination better and offers higher precision.

\subsection{RBO Photometry}
\begin{figure*}
    \centering
    \includegraphics[width=\linewidth]{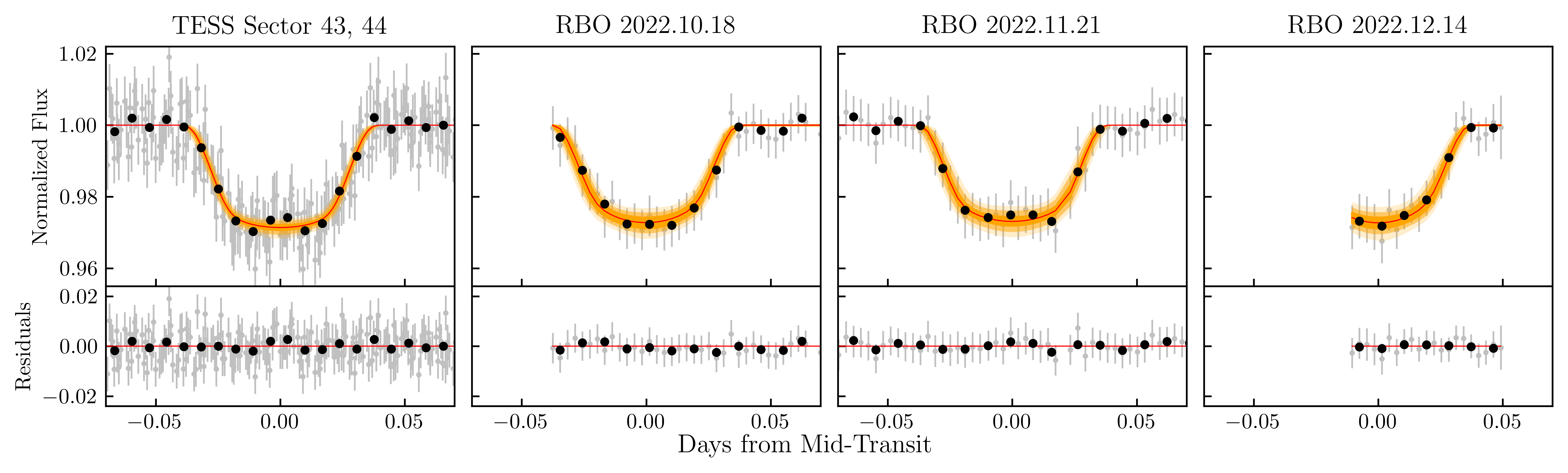}
    \caption{The raw (grey) and 10-minute-binned (black) phase-folded light curves of TESS and RBO photometry. The best joint fit model is shown in red, and the 1$\sigma$, 2$\sigma$, and 3$\sigma$ confidence intervals are shown in yellow with descending intensities. }
    \label{fig:photometry}
\end{figure*}

We used the Red Buttes Observatory 0.6 m telescope \citep[RBO;][]{Kasper2016_RBO} to observe three transits of TOI-5344~b during the nights of 2022 October 18, 2022 November 21, and 2022 December 14. The observations were carried out mildly defocused to a FWHM of $\sim 2.0\arcsec$, with exposure times of 240~s. The first two transits were observed in Bessell $I$ and the last in Bessell $R$, at airmasses of 1.07 to 2.03, 1.07 to 1.73, and 1.12 to 1.33, respectively. The raw data were reduced using a custom \texttt{python} differential aperture photometry pipeline based on the one outlined in \citet{Monson2017_TMMT}. The photometry is shown in Figure~\ref{fig:photometry}. 

\subsection{NESSI Speckle Imaging} \label{sec:nessi}
\begin{figure}
    \centering
    \includegraphics[width=\linewidth]{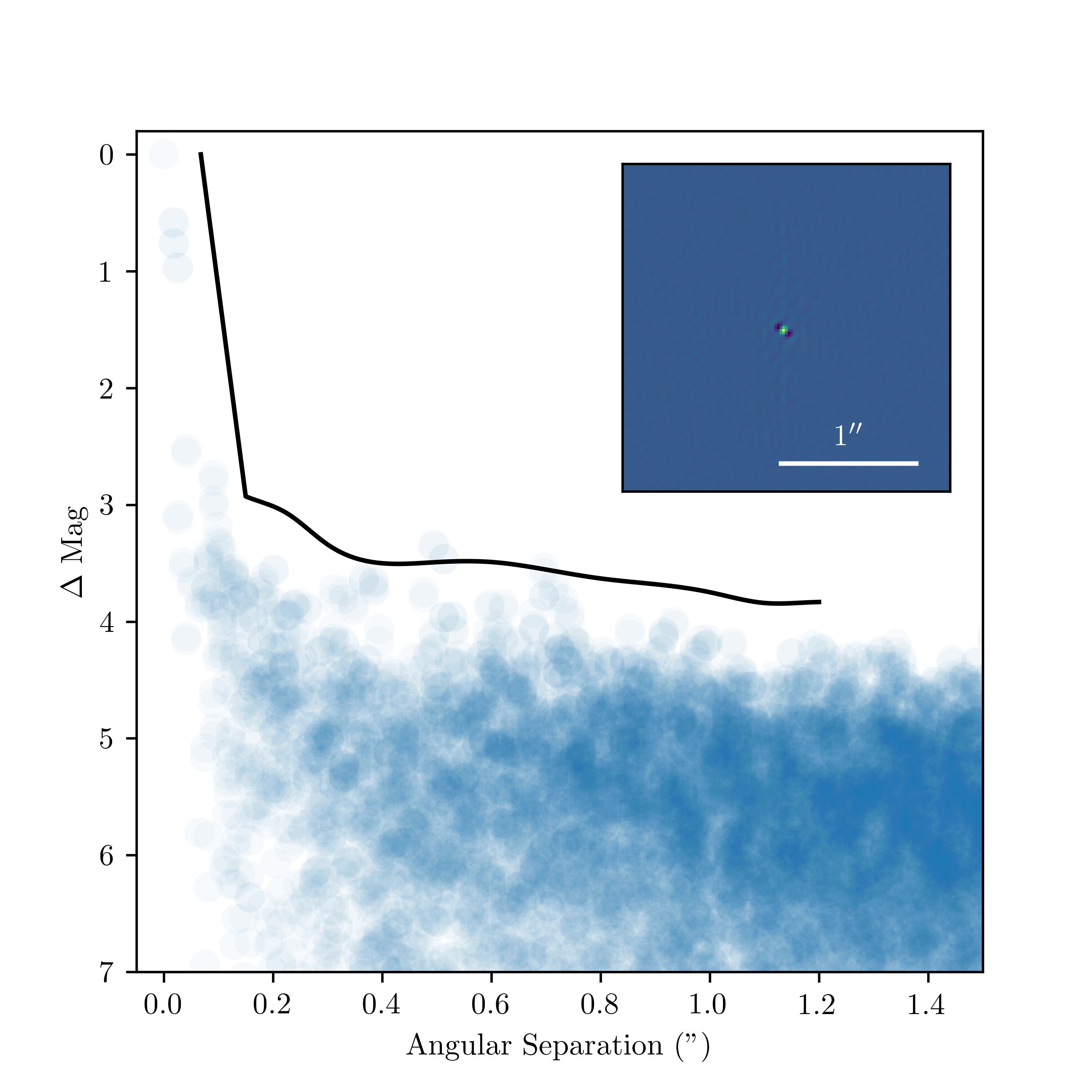}
    \caption{NESSI 5-$\sigma$ contrast curve of TOI-5344 in the Sloan $z'$ filter (black) with the reconstructed speckle image shown on top. The per-pixel magnitude difference is plotted in blue circles. No bright companions with $\Delta z' <$ 3.0 are observed between $0.\!\!\arcsec2$ and $1.\!\!\arcsec2$ of the target.}
    \label{fig:nessi}
\end{figure}

We observed TOI-5344 on the night of 2022 April 18 using the NN-Explore Exoplanet Stellar Speckle Imager \citep[NESSI;][]{NESSI} on the WIYN\footnote{The WIYN Observatory is a joint facility of the NSF’s National Optical-Infrared Astronomy Research Laboratory, Indiana University, the University of Wisconsin-Madison, Pennsylvania State University, the University of Missouri, the University of California-Irvine, and Purdue University.} 3.5 m telescope at Kitt Peak National Observatory to search for nearby companions or background sources. A sequence of 9000 40 ms images was taken in the Sloan $z'$ filter for a total integration time of 6 minutes. The reconstructed image \citep[produced according to][]{Howell11} and 5-$\sigma$ contrast curve are shown in Figure~\ref{fig:nessi}, which exclude nearby sources with magnitudes brighter than $\Delta z'$ = 3.0 at separations $>$ 0$.\!\!\arcsec2$.

\subsection{Habitable-zone Planet Finder Spectroscopy}
We observed TOI-5344 using the near-infrared Habitable-zone Planet Finder \citep[HPF;][]{HPF_1, HPF_2, optical_fiber} from 2022 September 11 to 2023 January 21. HPF is located at the 10 m Hobby-Eberly Telescope \citep[HET;][]{Ramsey98}. We processed the raw HPF data using \texttt{HxRGproc} algorithms \citep{HxRGproc} and calibrated the spectra wavelength by the method described in \cite{Stefansson20}. We then derived RV from the spectra using a modified version of the \texttt{SpEctrum Radial Velocity AnaLyser} pipeline \citep[SERVAL;][]{Zechmeister18}. Lastly, we performed the barycentric correction on the individual spectra with \texttt{barycorrpy}, a \texttt{python} implementation of the algorithms from \cite{Wright14} developed by \cite{Kanodia18}.

We obtained a total of 31 exposures on TOI-5344. The observations were conducted over 16 nights, with two exposures per visit taken on 15 of these nights. Two exposures were excluded from RV analysis due to poor weather conditions during the observations. The excluded exposures have S/N $\lesssim 23$ (per pixel at 1070 nm) and $\sigma \gtrsim 60$ m/s in comparison to median(S/N) $=46.37$ and median($\sigma$) $=27.21$ m/s for the remaining exposures. The selected 29 exposures of 969 s exposures are listed in Table~\ref{tab:RV}. 

\begin{figure*}
    \centering
    \includegraphics[width=\linewidth]{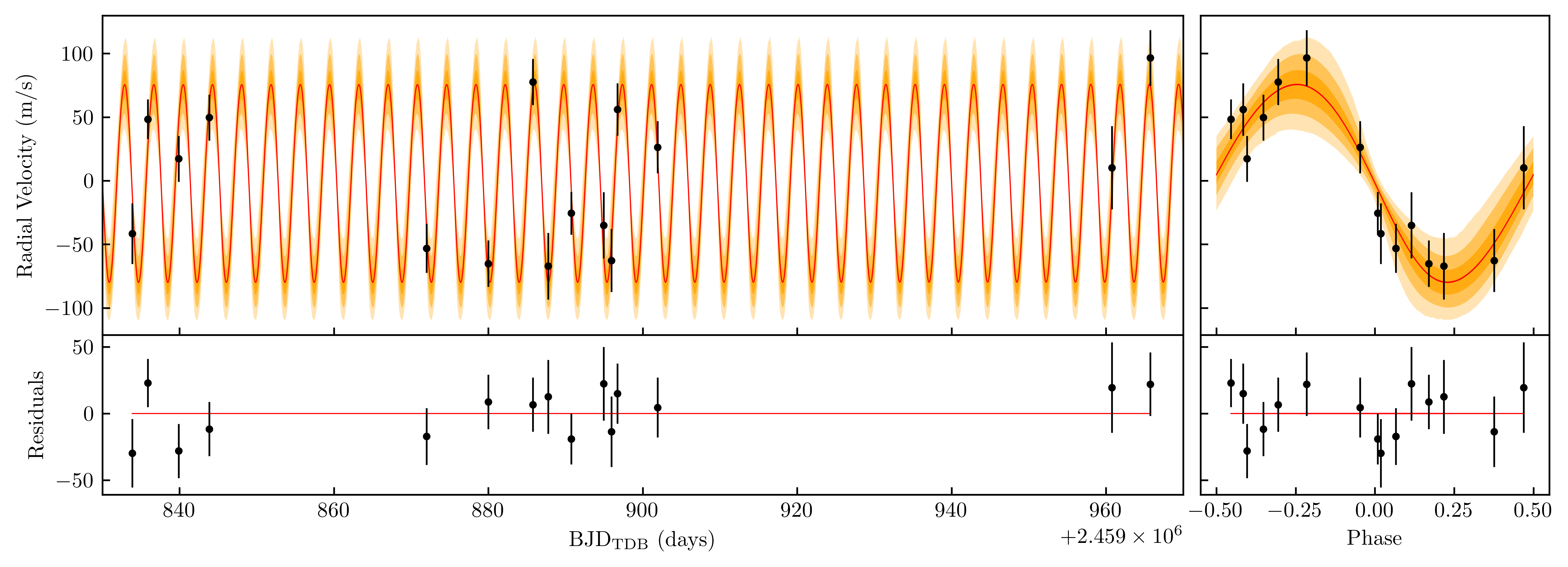}
    \caption{The binned HPF RV measurements of TOI-5344. Left: Time series of HPF RV binned by each night (black). Right: Phase-folded HPF RV (black). The best joint fit model is shown in red, and the 1$\sigma$, 2$\sigma$, and 3$\sigma$ confidence intervals are shown in yellow with descending intensities. }
    \label{fig:rv}
\end{figure*}

\begin{deluxetable}{cccc}
\tablecaption{TOI-5344 HPF RVs \label{tab:RV}}
\tablehead{\colhead{BJD} &\colhead{RV (m/s)} &\colhead{$\sigma$ (m/s)} & \colhead{S/N}}
\startdata
2459833.88521 & -140.64  & 35.66          & 37 \\
2459833.89605 & -128.20  & 32.16          & 42 \\
2459835.88467 & -54.65   & 21.32          & 59 \\
2459835.89622 & -31.54   & 22.73          & 55 \\
2459839.86792 & -85.36   & 24.20          & 52 \\
2459839.87947 & -61.81   & 27.13          & 48 \\
2459843.85663 & -37.43   & 25.96          & 49 \\
2459843.86833 & -47.27   & 25.11          & 50 \\
2459871.99008 & -142.65  & 27.29          & 46 \\
2459872.00165 & -147.97  & 27.21          & 47 \\
2459879.96989 & -188.15  & 24.67          & 50 \\
2459879.98144 & -120.23  & 27.08          & 46 \\
2459885.75243 & 10.10    & 25.08          & 49 \\
2459885.76427 & -41.42   & 26.21          & 49 \\
2459887.73459 & -111.64  & 38.91          & 34 \\
2459887.74608 & -198.60  & 35.35          & 37 \\
2459890.73701 & -113.92  & 22.71          & 55 \\
2459890.74852 & -122.27  & 24.76          & 50 \\
2459894.93751 & -127.13  & 26.05          & 48 \\
2459895.92305 & -174.35  & 35.85          & 36 \\
2459895.93450 & -137.26  & 34.17          & 37 \\
2459896.71145 & -49.23   & 29.10          & 44 \\
2459896.72291 & -23.17   & 29.10          & 43 \\
2459901.90475 & -84.61   & 26.59          & 46 \\
2459901.91588 & -38.49   & 32.08          & 39 \\
2459960.74994 & -84.32   & 40.35          & 33 \\
2459960.76167 & -77.37   & 56.03          & 25 \\
2459965.73717 & -3.14    & 27.93          & 45 \\
2459965.74850 & 16.40    & 35.54          & 36 \\
\enddata
\textbf{Notes.}
\tablenotetext{}{S/N is calculated per pixel at 1070 nm. All observations have exposure times of 969 s. }
\end{deluxetable}

\section{Stellar Parameters}  \label{sec:stellar}
\subsection{HPF-SpecMatch} \label{sec:hpf_specmatch}

We used the \texttt{HPF-SpecMatch} package \citep{Stefansson20} to estimate stellar parameters from HPF spectra using a two-step $\chi^2$-based algorithm. It identifies the best-matching library stars for the target spectrum, with a library of 166 stars spanning 2700 K $< T_{\text{eff}} < $ 6000 K, $4.3<\log g_{\star}<5.3$, and -0.5 $\lesssim$ [Fe/H] $\lesssim$ 0.5. The target spectrum is compared to all library spectra, and only the top five best-fit stars are used to generate a composite spectrum that closely matches the target. 

For TOI-5344, HPF order index 5 (8534-8645 \AA) was used for spectral matching because it has the least telluric contamination. No strong rotational broadening is observed in the spectral match, resulting in an upper limit of $v \sin i < 2$ km s$^{-1}$ for TOI-5344. This is consistent with TESS photometry, which detects no significant rotational modulation. Further discussion on the absence of rotation signal in long baseline photometry is in Section \ref{sec:rotation}. We list the derived effective temperature, metallicity, and surface gravity of TOI-5344 in Table \ref{tab:stellar}, which agree within 1$\sigma$ of the results from \cite{hartman23}.

\begin{figure*}
    \centering
    \includegraphics[width=\textwidth]{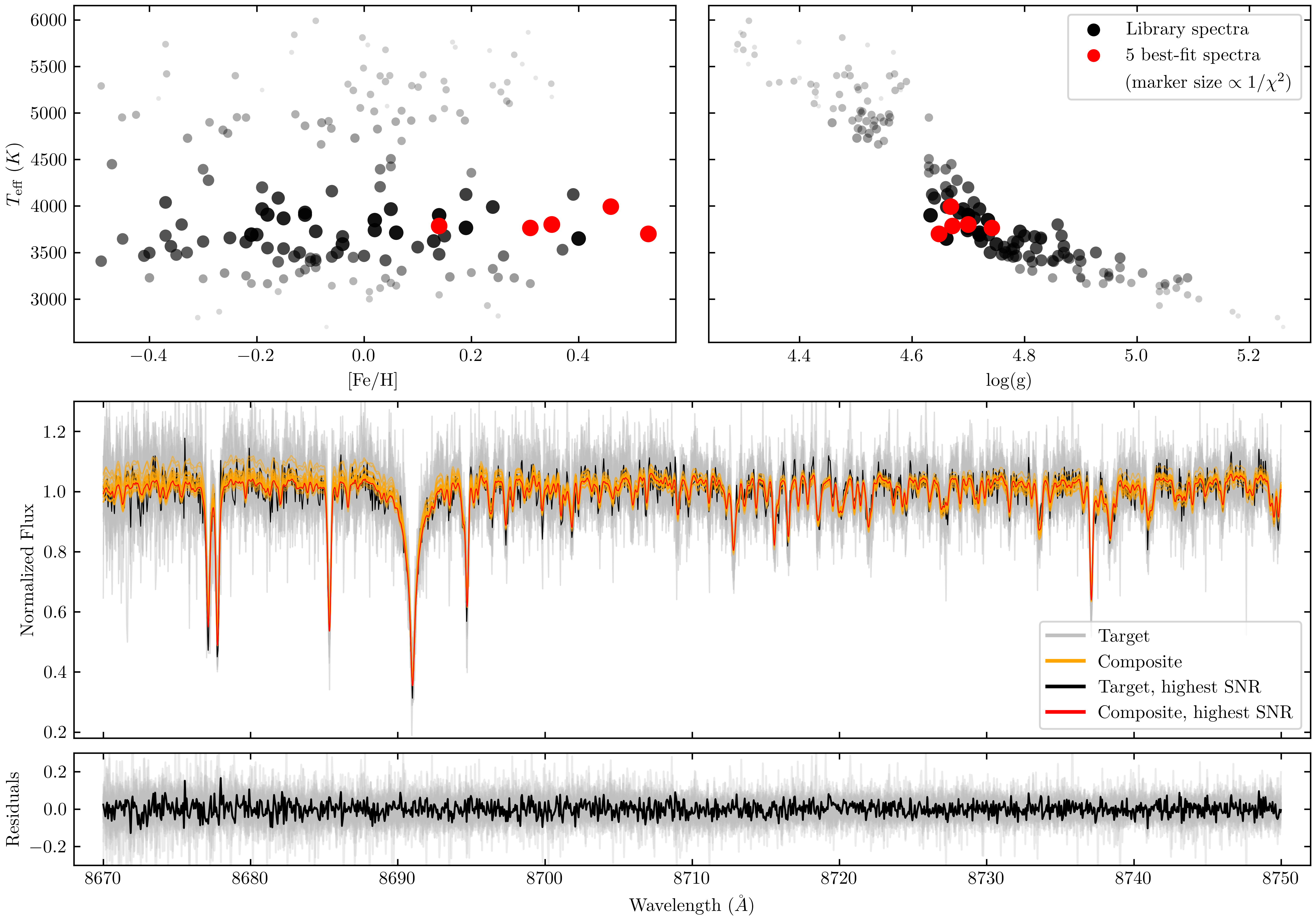}
    \caption{The \texttt{HPF-SpecMatch} spectra fit results for order index 5. Top: Two plots showing the five best-fit stars (red) selected to compose the spectra. The size and transparency of these five stars and the other library stars (black) are inversely proportional to the calculated $\chi^2$ initial value where we compare TOI-5344 spectra to the library star spectra. Larger and darker points signify a lower $\chi^2$ initial value, implying a better fit to the target spectrum. Middle: All HPF order index 5 spectra of TOI-5344 (grey) and their best-fit composite (yellow). The highest S/N spectrum is shown in black and its best-fit composite is in red. Bottom: The residuals of the best-fit composites for all spectra (grey) and for the highest S/N spectrum (black). The featureless residuals suggest a high-quality fit.}
    \label{fig:hpfspecmatch}
\end{figure*}

Notably, the metallicity of [Fe/H] = $0.48 \pm 0.12$ is close to the \texttt{HPF-SpecMatch} library's upper limit, raising a question of whether TOI-5344 has a metallicity higher than 0.5. However, despite the fact that the five best-fit stars (GJ 205, GJ 134, GJ 96, BD+29 2279, and GJ 895) are among the highest metallicity in the library, we formally adopt the metallicity returned by \texttt{HPF-SpecMatch} given the featureless residual displayed in Figure~\ref{fig:hpfspecmatch}, where it could be potentially biased to be too low given the boundary conditions imposed by the library stars. We also estimated the metallicity with a photometric calibration using METaMorPHosis \citep{Duque-Arribas22}, which yields [Fe/H] = 0.35 $\pm$ 0.13. While the uncertainties are large, both metallicity determination methods agree that the [Fe/H] of TOI-5344 is considerably higher than Solar, at a 3-4 $\sigma$ significance level. \cite{hartman23} reports a comparable spectroscopic metallicity of [Fe/H] = $0.390 \pm 0.090$, confirming TOI-5344's super-Solar metallicity. 

\begin{deluxetable*}{lccr}
\tablewidth{0pt}
\tablecaption{Summary of Stellar Parameters for TOI-5344 \label{tab:stellar}}
\tablehead{\colhead{\hspace{.3cm}Parameter} &\colhead{\hspace{1.5cm}Description}\hspace{1.5cm} &\colhead{Value} & \colhead{Reference}}
\startdata
\multicolumn{4}{l}{Main identifiers:} \\
\hspace{.3cm}TOI & TESS Object of Interest & 5344 & TESS mission \\
\hspace{.3cm}TIC & TESS Input Catalogue & 16005254 & Stassun \\
\hspace{.3cm}2MASS & ... & J04130384+2054550 & 2MASS \\
\hspace{.3cm}Gaia DR3 & ... & 52359538285081728 & Gaia DR3 \\ 
\multicolumn{4}{l}{\text{Equatorial Coordinates, Proper Motion and Spectral Type:}} \\
\hspace{.3cm}$\alpha_{\text{J2015.5}}$ & Right Ascension (RA) & 63.266 $\pm$ 0.021 & Gaia DR3 \\
\hspace{.3cm}$\delta_{\text{J2015.5}}$ & Declination (Dec) & 20.915 $\pm$ 0.013 & Gaia DR3 \\
\hspace{.3cm}$\mu_\alpha$ & Proper motion (RA, mas/yr) & 40.32 $\pm$ 0.03 & Gaia DR3 \\
\hspace{.3cm}$\mu_\delta$ & Proper motion (Dec, mas/yr) & -22.19 $\pm$ 0.02 & Gaia DR3 \\
\hspace{.3cm}$\varpi$ & Parallax (mas) & 7.31 ± 0.02 & Gaia DR3 \\
\hspace{.3cm}$d$ & Distance (pc) & 136.1 $\pm$ 0.5 & Bailer-Jones \\
\hspace{.3cm}$A_{V,max}$ & Maximum visual extinction & 0.012 & Green \\
\multicolumn{4}{l}{\text{Optical and Near-infrared Magnitudes:}} \\
\hspace{.3cm}$B$ & Johnson B mag & 16.888 $\pm$ 0.131 & APASS \\
\hspace{.3cm}$V$ & Johnson V mag & 15.288 $\pm$ 0.066 & APASS \\
\hspace{.3cm}$g'$ & Sloan $g'$ mag & 16.09 $\pm$ 0.069 & APASS \\
\hspace{.3cm}$r'$ & Sloan $r'$ mag & 14.643 $\pm$ 0.055 & APASS \\
\hspace{.3cm}$i'$ & Sloan $i'$ mag & 13.665 $\pm$ 0.083 & APASS \\
\hspace{.3cm}$J$ & $J$ mag & 11.799 $\pm$ 0.021 & 2MASS \\
\hspace{.3cm}$H$ & $H$ mag & 11.087 $\pm$ 0.022 & 2MASS \\
\hspace{.3cm}$K_s$ & $K_s$ mag & 10.86 $\pm$ 0.018 & 2MASS \\
\hspace{.3cm}$W1$ & WISE1 mag & 10.739 $\pm$ 0.024 & WISE \\
\hspace{.3cm}$W2$ & WISE2 mag & 10.728 $\pm$ 0.02 & WISE \\
\hspace{.3cm}$W3$ & WISE3 mag & 10.554 $\pm$ 0.113 & WISE \\
\multicolumn{4}{l}{\text{Spectroscopic Parameters:}} \\
\hspace{.3cm}$T_{\text{eff}}$ & Effective temperature (K) & $3770 \pm 88$ & This work \\
\hspace{.3cm}$\text{[Fe/H]}$ & Metallicity (dex) & $0.48 \pm 0.12$ & This work \\
\hspace{.3cm}$\log g_{\star}$ & Surface gravity (cgs units) & $4.68 \pm 0.05$ & This work \\
\multicolumn{4}{l}{\text{Model-Dependent Stellar SED and Isochrone fit Parameters:}} \\
\hspace{.3cm}$M_*$ & Mass ($\Msun$) & $0.59^{+0.02}_{-0.03}$ & This work \\
\hspace{.3cm}$R_*$ & Radius ($\Rsun$) & $0.563 \pm 0.016$ & This work \\
\hspace{.3cm}$L_*$ & Luminosity ($L_\odot$) & $0.0570^{+0.0017}_{-0.0015}$ & This work \\
\hspace{.3cm}$\rho_*$ & Density (g cm$^{-3}$) & $4.7 \pm 0.3$ & This work \\
\hspace{.3cm}Age & Age (Gyrs) & $7.1 \pm 4.5$ & This work \\
\hspace{.3cm}$A_v$ & Visual extinction (mag) & $0.006\pm 0.004$ & This work \\
\multicolumn{4}{l}{\text{Other Stellar Parameters:}} \\
\hspace{.3cm}$v \sin i $ & Rotational velocity (km s$^{-1}$) & $<2 $ & This work \\
\hspace{.3cm}$\Delta$ RV & “Absolute” radial velocity (km s$^{-1}$) & $46.78 \pm 0.10 $ & This work \\
\hspace{.3cm}$U, V, W$ & Galactic velocities (km s$^{-1}$) & $-49.86 \pm 0.09, -22.86 \pm 0.08, -8.29 \pm 0.05$ & This work \\
\hspace{.3cm}$U, V, W^a$ & Galactic velocities in LSR (km s$^{-1}$) & $-38.76 \pm 0.84, -10.62 \pm 0.55, -1.04 \pm 0.42$ & This work
\enddata
\vspace{0.1cm}
\textbf{Notes.} \\
References: Stassun \citep{Stassun18}, 2MASS \citep{Cutri03}, Gaia EDR3 \citep{Gaia, Gaia_DR3}, Bailer-Jones \citep{Bailer-Jones18}, Green \citep{Green19}, American Association of Variable Star Observers Photometric All Sky Survey \citep[APASS;][]{APASS}, Wide-field Infrared Survey Explorer \citep[WISE;][]{Wright10}.
\tablenotetext{a}{The barycentric UVW velocities are converted into the local standard of rest (LSR) velocities using the constants from \cite{Schonrich10}.}
\end{deluxetable*}

\subsection{Model-Dependent Stellar Parameters} \label{sec:exofast}
\begin{figure}
    \centering
    \includegraphics[width=\linewidth]{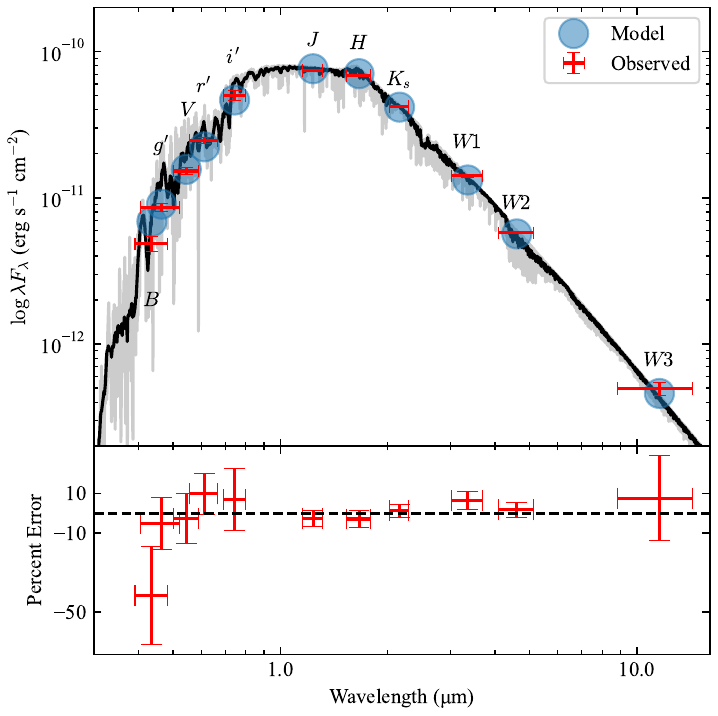}
    \caption{SED of TOI-5344 showing the broadband photometric measurements from Table \ref{tab:stellar} in red (the $x$-errorbar is the bandpass width) and the derived MIST model fluxes (as blue circles). A NextGen BT-SETTL spectrum \citep{Allard12} is overlaid for reference in gray (a smoothed version in black); the model spectrum is not used when fitting the SED.}
    \label{fig:sed}
\end{figure}

We used the \texttt{EXOFASTv2} package \citep{exofastv2} to model the spectral energy distribution (SED) of TOI-5344, which allows us to obtain the model-dependent stellar parameters including stellar mass, radius, luminosity, and age. We chose the default Modules for Experiments in Stellar Astrophysics Isochrones and Stellar Tracks (MIST) model grids \citep{Dotter16, Choi16} and placed Gaussian priors on the reliable broadband photometry from APASS, 2MASS, and WISE, as well as on the derived spectroscopic stellar parameters from \texttt{HPF-SpecMatch} and the extinction-corrected geometric distance from \cite{Bailer-Jones21}. We show the SED fit in Figure \ref{fig:sed} and list the derived model-dependent stellar parameters in Table~\ref{tab:stellar}. Our adopted stellar parameters derived from \texttt{EXOFASTv2} agree within 1$\sigma$ of the results from \cite{hartman23}, including the stellar mass, radius, luminosity, and age. We omit the \texttt{EXOFASTv2} estimations of the effective temperature, metallicity, and surface gravity since they are statistically equivalent to the adopted \texttt{HPF-SpecMatch} results. 

\subsection{Galactic Kinematics}
We calculated the UVW velocities in the barycentric frame using \texttt{GALPY} \citep{Bovy15} based on the systemic velocity from HPF and proper motion from Gaia DR3 \citep{Gaia_DR3}. These velocities, along with the ones in the local standard of rest using offsets from \cite{Schonrich10}, are provided in Table 3. We classify TOI-5344 as a field star in the thin disk with a $99.9\%$ probability based on its systemic velocity, position, and proper motion using the BANYAN $\Sigma$ tool \citep{Gagne18}. We further support this conclusion by referring to Equation A.1 from \cite{Bensby14}, which indicates a 72 times higher likelihood for TOI-5344 to belong to the thin disk rather than the thick disk using the UVW velocities. 

\subsection{No Detectable Stellar Rotation Signal \label{sec:rotation}}
\begin{figure}
    \centering
    \includegraphics[width=\linewidth]{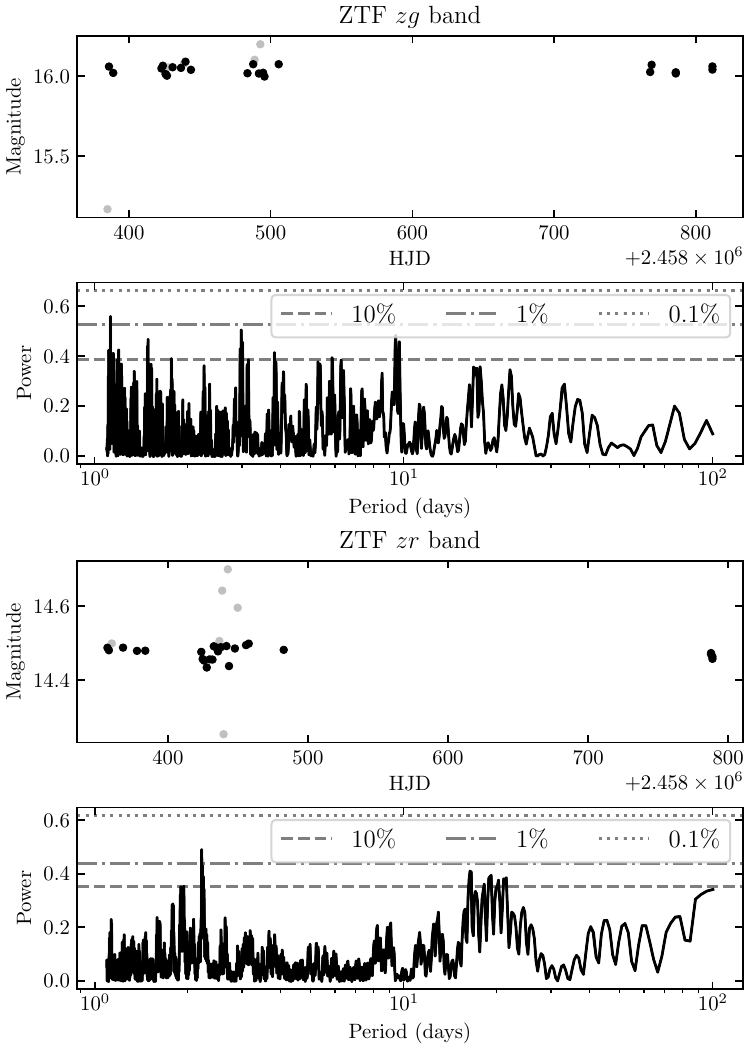}
    \caption{The ZTF $zg$ and $zr$ band photometries and their GLS periodograms of TOI-5344. Grey points are discarded in the analysis since they have low quality flags. Despite several peaks reaching $\sim$1\% FAP, their significance is limited by the small sample size. }
    \label{fig:ZTF}
\end{figure}
\begin{figure}
    \centering
    \includegraphics[width=\linewidth]{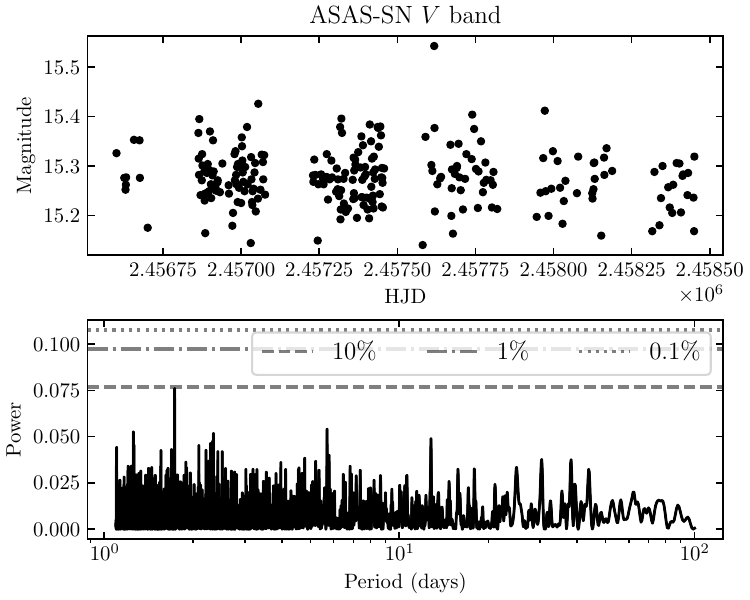}
    \caption{The ASAS-SN $V$ band photometry and its GLS periodogram of TOI-5344. The false alarm probability (FAP) is shown in dashed lines. No significant stellar variation signal is observed in the GLS. }
    \label{fig:asassn}
\end{figure}

We performed the Generalized Lomb-Scargle periodogram \citep[GLS periodogram;][]{Lomb76, Scargle82, Zechmeister09} analysis on long baseline public photometric data to search for stellar rotation signals. TOI-5344 was observed by Zwicky Transient Facility \citep[ZTF;][]{Masci19} over $\sim 400$ days (2018 August 26 - 2019 November 24) in $zg$ (27 observations) and $zr$ (31 observations) bands, and by the All-Sky Automated Survey for Supernovae \citep[ASAS-SN;][]{Kochanek17} for a span of $\sim 2000$ days (2013 November 4 - 2018 November 28) with only $V$ band (244 observations). No significant stellar rotation signal is detected in the $zg$ or $zr$ bands of ZTF\footnote{Data acquired from ZTF DR18: \url{https://www.ztf.caltech.edu/ztf-public-releases.html}} (Figure \ref{fig:ZTF}), nor the $V$ band of ASAS-SN (Figure \ref{fig:asassn}). Despite several peaks of both ZTF GLS periodograms reaching $\sim$1\% false alarm probability (FAP), they are not considered significant given the small sample sizes of only $\sim 20$ epochs for each band. The GLS periodogram of the TESS light curve also shows no remarkable signal. 

The absence of photometric rotational modulation over the long observation baseline indicates an inactive star, which is supported by the lack of emission or any detectable temporal changes in the cores of the Calcium II NIR triplet \citep{Mallik97,Cincunegui07,Robertson16,Martin17} in the HPF spectra. Recalling a $v \sin i$ upper limit of 2 km s$^{-1}$ (Section~\ref{sec:hpf_specmatch}) and an age estimate of 7.1 $\pm$ 4.5 Gyr (Section~\ref{sec:exofast}), we conclude that TOI-5344 is an inactive and likely old star with a slow rotation rate. 

\subsection{Ruling out Stellar Companions}
TOI-5344 is in a relatively sparse field according to Gaia (Figure~\ref{fig:tess}). The closest Gaia neighbor is TIC 16005253 (Gaia DR3 52359533989348224), which is $\sim 21$ arcsecs away and $\sim 3$ magnitudes dimmer. No co-moving nearby stars are observed based on Gaia DR3 proper motions. Gaia reports a Renormalised Unit Weight Error (RUWE) of $\sim$1.035 for TOI-5344, indicating a low possibility for an unresolved companion. We further confirm the lack of stellar companions with magnitudes brighter than $\Delta r'$ = 3.0 at separations $>$ 0$.\!\!\arcsec3$ using NESSI in Section~\ref{sec:nessi}. 


\section{Joint Fitting of Photometry and RVs} \label{sec:joint_fit}

We performed a joint fit of photometry and HPF RVs (binned by night) using \textsf{exoplanet} \citep{exoplanet}, which is based on \texttt{PyMC3}, the Hamiltonian Monte Carlo package (Salvatier et al. 2016). The \textsf{exoplanet} package utilizes \texttt{starry} \citep{exoplanet:luger18, exoplanet:agol20} to simulate the planetary transits. It employs the analytical transit models by \cite{Mandel02} and a quadratic limb-darkening law. The limb-darkening priors are implemented in \textsf{exoplanet} following the reparameterization by \cite{Kipping13} for uninformative sampling. Each transit (Figure~\ref{fig:photometry}) is fitted independently with specific limb-darkening coefficients. We experimented with a Gaussian Process (GP) kernel for TESS, which shows non-periodic correlated noise. We found consistent results for the planetary radius regardless of whether we used a GP, so we adopted a model that does not utilize the GP for simplicity. 

Despite the aforementioned closest contamination source (TIC 16005247, Figure~\ref{fig:tess}) being $\sim 21$ arcsecs away, it still strongly contaminates the TOI-5344~b TESS light curve since its centroid is in an adjacent TESS pixel. We decided to let the TESS light curve dilution factor vary even with TGLC's decontamination capability. The true transit depth was derived from 3 RBO transit observations (Figure~\ref{fig:photometry}), which should have no contamination source with $\Delta z'$ < 3.0 at separations $>$ 0$.\!\!\arcsec2$ based on the NESSI analysis (Section \ref{sec:nessi}). The $I$ and $R$ bands of RBO show no evidence of inconsistent transit depth or chromaticity. The joint fit gives $D_{\text{TGLC}}$ = 1.075$^{+0.068}_{-0.065}$, indicating that TGLC's decontamination process is consistent within the photometric precision\footnote{See Section \ref{sec:observations} for the explanation of $D_{\text{TGLC}}$ > 1. }.

The best-fit transit model is shown in a phase-folded plot in Figure~\ref{fig:photometry}, and the best-fit RV model is plotted in Figure~\ref{fig:rv} both as a time-series and phase-folded. A summary of the inferred system parameters along with their corresponding confidence intervals is presented in Table~\ref{tab:planet}. TOI-5344~b is a Saturn-like planet with a radius of $9.7 \pm 0.5 \ \Rearth$ ($0.87 \pm 0.04 \ \Rjup$), a mass of 135$^{+17}_{-18} \ \Mearth$ (0.42$^{+0.05}_{-0.06} \ \Mjup$), and a density of 0.80$^{+0.17}_{-0.15}$ g cm$^{-3}$. 

All planetary parameters in Table~\ref{tab:planet} agree within 1$\sigma$ with the corresponding values from \cite{hartman23}, except for the planet radius. \cite{hartman23} reports a planet radius of $0.946 \pm 0.021 \ \Rjup$, which is larger than our calculation of $0.87 \pm 0.04 \ \Rjup$ by more than 1$\sigma$. Both their calculation of $R_p/R_*$ (0.1653 $\pm$ 0.0014) and stellar radius estimate (0.588 $\pm$ 0.011 $\Msun$) are larger than our values (0.159$^{+0.006}_{-0.005}$ and 0.563 $\pm$ 0.016 $\Msun$, respectively), but both fall within 1$\sigma$. The radius discrepancy arises from variations in stellar models and adopted photometries, without significantly altering the interpretation of the planet's nature. 

\begin{deluxetable*}{lccc}
\tablewidth{0pt}
\tablecaption{Summary of Planetary Parameters for TOI-5344~b \label{tab:planet}}
\tablehead{\colhead{Parameter} &\colhead{Description} &\colhead{Prior} &\colhead{Value$^a$}}
\startdata
\multicolumn{4}{l}{Orbital Parameters:} \\
\hspace{.3cm}$P$ & Orbital Period (days) & $\mathcal{N}(3.79, 0.1)$ & 3.792622$^{+0.000010}_{-0.000010}$ \\
\hspace{.3cm}$e$ & Eccentricity & $\sqrt{e} \cos \omega$: $\mathcal{U}(-1, 1)$ & < 0.21$^b$, 0.06$^{+0.07}_{-0.04}$ \\
\hspace{.3cm}$\omega$ & Argument of Periastron (degrees) & $\sqrt{e} \sin \omega$: $\mathcal{U}(-1, 1)$ & 95$^{+53}_{-214}$ \\
\hspace{.3cm}$K$ & Semi-amplitude Velocity (m s$^{-1}$) & ... & $78 \pm 10$ \\
\hspace{.3cm}$\gamma_{\text{HPF}}$ & Systemic Velocity$^c$ (m s$^{-1}$) & $\mathcal{N}(-82, 20000)$ & $-92^{+6}_{-7}$ \\
\hspace{.3cm}$dv/dt$ & RV trend (m s$^{-1}$ yr$^{-1}$) & $\mathcal{N}(0, 5)$ & $1 \pm 5$ \\
\hspace{.3cm}$\sigma_{\text{HPF}}$ & RV jitter (m s$^{-1}$) & $\mathcal{U}(0.001,1000)$ & 10$^{+8}_{-7}$ \\
\multicolumn{4}{l}{Transit Parameters:} \\
\hspace{.3cm}$T_C$ & Transit Midpoint (BJD$_{\text{TDB}}$) & $\mathcal{N}(2459477.313, 0.01)$ & $2459477.3131 \pm {0.0007}$ \\
\hspace{.3cm}$R_p/R_*$ & Scaled Radius & $\mathcal{LN}(-1.8, 1)$ & 0.159$^{+0.005}_{-0.006}$ \\
\hspace{.3cm}$a/R_*$ & Scaled Semimajor Axis & ... & 15.3$^{+0.5}_{-0.4}$ \\
\hspace{.3cm}$i$ & Orbital Inclination (degrees) & ... & $87.5 \pm 0.2$ \\
\hspace{.3cm}$b$ & Impact Parameter & $\mathcal{U}(0,1)$ & 0.66$^{+0.05}_{-0.08}$ \\
\hspace{.3cm}$T_{14}$ & Transit Duration (hours) & ... & 1.81$^{+0.10}_{-0.07}$ \\
\hspace{.3cm}$\sigma_{\text{TESS}}$ & Photometric Jitter$^d$ (ppm) & $\mathcal{LN}(-5, 2)$ & 158$^{+150}_{-98}$ \\
\hspace{.3cm}$\sigma_{\text{RBO 20221018}}$ & Photometric Jitter$^d$ (ppm) & $\mathcal{LN}(-5, 2)$ & 464$^{+503}_{-304}$ \\
\hspace{.3cm}$\sigma_{\text{RBO 20221121}}$ & Photometric Jitter$^d$ (ppm) & $\mathcal{LN}(-5, 2)$ & 549$^{+620}_{-364}$ \\
\hspace{.3cm}$\sigma_{\text{RBO 20221214}}$ & Photometric Jitter$^d$ (ppm) & $\mathcal{LN}(-4, 2)$ & 771$^{+974}_{-531}$ \\
\hspace{.3cm}$D_{\text{TGLC}}$ & Dilution$^e$ & $\mathcal{U}(0.1,1.5)$ & 1.075$^{+0.068}_{-0.065}$ \\
\multicolumn{4}{l}{Stellar Parameters} \\
\hspace{.3cm}$M_*$ & Mass ($\Msun$) & $\mathcal{BN}(0.591, 0.02, 0, 1.5)$ & $0.59 \pm 0.02$ \\
\hspace{.3cm}$R_*$ & Radius ($\Rsun$) & $\mathcal{BN}(0.563, 0.02, 0, 1.5)$ & $0.56 \pm 0.02$ \\
\hspace{.3cm}$T_{\text{eff}}$ & Effective temperature (K) & $\mathcal{BN}(3757, 51, 2000, 7000)$ & $3757 \pm 50$ \\
\hspace{.3cm}$u_1^f$ & Limb-darkening parameter & $\mathcal{U}(0,1)$ & ... \\
\hspace{.3cm}$u_1^f$ & Limb-darkening parameter & $\mathcal{U}(0,1)$ & ... \\
\multicolumn{4}{l}{Planetary Parameters:} \\
\hspace{.3cm}$M_p$ & Mass (M$_\oplus$) & $\mathcal{U}(0.1,3 \times 10^6)$ & $135^{+17}_{-18}$ \\
 & Mass (M$_{\text{J}}$) & ... & $0.42^{+0.05}_{-0.06}$ \\
\hspace{.3cm}$R_p$ & Radius (R$_\oplus$) & ... & $9.7 \pm 0.5$ \\
 & Radius (R$_{\text{J}}$) & ... & $0.87 \pm 0.04$ \\
\hspace{.3cm}$\rho_p$ & Density (g cm$^{-3}$) & ... & 0.80$^{+0.17}_{-0.15}$ \\
\hspace{.3cm}$a$ & Semimajor Axis (AU) & ... & $0.0400 \pm 0.0005$ \\
\hspace{.3cm}$\langle F \rangle$ & Average Incident Flux$^g$ ($10^5$ W m$^{-2}$) & ... & $0.48 \pm 0.04$ \\
\hspace{.3cm}$S$ & Planetary Insolation (S$_\oplus$) & ... & $35 \pm 3$ \\
\hspace{.3cm}$T_{\text{eq}}$ & Equilibrium Temperature$^h$ (K) & ... & $679 \pm 14$
\enddata
\vspace{0.1cm}
\textbf{Notes.} \\
Normal prior: $\mathcal{N}$(mean, standard deviation); Uniform prior: $\mathcal{U}$(lower, upper); Log-normal prior: $\mathcal{LN}$(mean, standard deviation); Bounded normal prior: $\mathcal{BN}$(mean, standard deviation, lower, upper)
\tablenotetext{a}{The reported values refer to the 16\%–50\%–84\% percentile of the posteriors.}
\tablenotetext{b}{Eccentricity upper limit, 95\% confidence. }
\tablenotetext{c}{In addition to the “Absolute” radial velocity from Table~\ref{tab:stellar}.}
\tablenotetext{d}{Jitter (per observation) added in quadrature to photometric instrument error.}
\tablenotetext{e}{TGLC dilution is set free, using ground-based photometry as the dilution-free reference. }
\tablenotetext{f}{Same Limb-darkening priors are applied to each photometry separately. Values are omitted. }
\tablenotetext{g}{We use the solar flux constant (1360.8 W m$^{-2}$) to convert insolation to incident flux. }
\tablenotetext{h}{We assume the planet to be a blackbody with zero albedo and perfect energy redistribution to estimate the equilibrium temperature.}

\end{deluxetable*}

\section{Discussion} \label{sec:discussion}
\subsection{TOI-5344~b in the GEMS parameter space} \label{sec:params}

\begin{figure*}
    \centering
    \includegraphics[width=\textwidth]{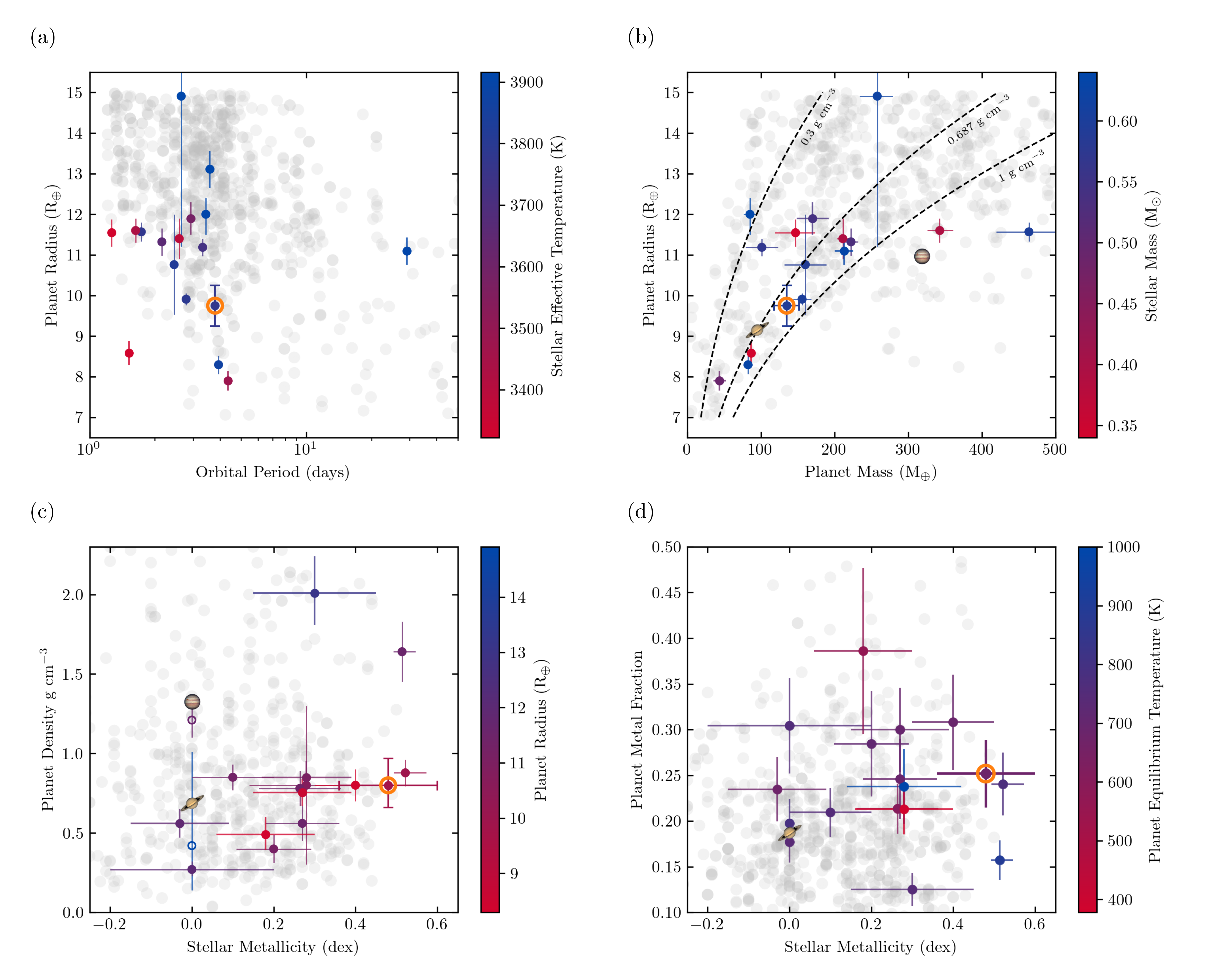}
    \caption{TOI-5344~b (highlighted with orange circle) in the transiting GEMS parameter spaces. All 17 Transiting GEMS are color-coded respectively, and transiting giants around FGK-dwarf stars are shown in grey circles. Panel (a) shows the Period-Radius plane. Panel (b) shows the Mass-Radius plane including Jupiter and Saturn as comparisons. TOI-4201~b \citep{Delamer23} has a planet mass larger than $800\ \Mearth$ and is not shown here. Three density lines are plotted, including Saturn's density ($0.687$ g cm$^{-3}$). We observe that TOI-5344~b is close to Saturn in this plane. Panel (c) shows the Metallicity-Density plane, where we note a lack of metal-poor host stars for transiting GEMS. The two empty circles represent two systems discarded in the statistics because of poor metallicity constraints. Panel (d) shows the Metallicity-Metal Fraction plane, and all 17 transiting GEMS are calculated to have a metal mass fraction of more than 10\% based on \cite{Thorngren16}. 
    }
    \label{fig:planet_params}
\end{figure*}

Figure~\ref{fig:planet_params} shows TOI-5344~b in the parameter space of all transiting GEMS. We draw our sample from the NASA Exoplanet Archive \citep[NEA;][]{Akeson13} and add two recent GEMS examples: TOI-4201 b \citep{Delamer23, Gan23_4201, hartman23} and TOI-4860~b \citep{almenara23, Triaud23}. In all four panels, we restrict our sample to planets with radii 8 $\Rearth$ $\lesssim R_p \lesssim$ 15 $\Rearth$ with host star effective temperatures $T_{\text{eff}} <$ 4000 K \citep[the M dwarf effective temperature scale,][]{Rajpurohit13, Baraffe15}. Only the planets with $\geq 3\sigma$ mass measurements are selected, resulting in a final sample size of 17 transiting GEMS (Table \ref{tab:gems}). The transiting giant planets around FGK-dwarf stars are plotted in the background following the same mass cut. 

\begin{deluxetable*}{lcccccc}
\tablewidth{0pt}
\tablecaption{Transiting GEMS \label{tab:gems}}
\tablehead{
    \colhead{Planet} & \colhead{Radius} & \colhead{Mass} & \colhead{Density} & \colhead{[Fe/H]} & \colhead{Metal Mass$^a$} & \colhead{Reference} \\ [-1.5ex]
    \colhead{} & \colhead{($\Rearth$)} & \colhead{($\Mearth$)} & \colhead{(g cm$^{-3}$)} & \colhead{dex} & \colhead{($\Mearth$)} & \colhead{}}
\startdata
TOI-3984 A b & 7.9$\pm 0.2$ & 44$^{+9}_{-8}$ & 0.49$^{+0.11}_{-0.10}$ & 0.18$^{+0.12}_{-0.12}$  & 37$^{+5}_{-5}$ & \cite{Canas23} \\
TOI-3629 b & 8.3$^{+0.2}_{-0.2}$ & 83$^{+6}_{-6}$ & 0.8$^{+0.1}_{-0.1}$ & 0.4$^{+0.1}_{-0.1}$  & 25$^{+4}_{-4}$ & \cite{Canas22} \\
TOI-4860 b & 8.6$^{+0.3}_{-0.3}$ & 86.7$^{+1.9}_{-1.9}$ & 0.75$^{+0.09}_{-0.09}$ & 0.27$^{+0.12}_{-0.12}$  & 26$^{+4}_{-4}$& \cite{almenara23} \\
TOI-5344~b & $9.7^{+0.5}_{-0.5}$ & $135^{+17}_{-18}$ & $0.80^{+0.17}_{-0.15}$ & 0.48$^{+0.12}_{-0.12}$  & $34^{+6}_{-6}$ & This work \\ 
HATS-75 b & 9.91$^{+0.15}_{-0.15}$ & 156$^{+12}_{-12}$ & 0.88$^{+0.08}_{-0.08}$ & 0.52$^{+0.05}_{-0.03}$  $^b$& 38$^{+5}_{-5}$ & \cite{Jordan22} \\
Kepler-45 b & 10.8$^{+1.2}_{-1.2}$ & 160$^{+29}_{-29}$ & 0.8$^{+0.5}_{-0.5}$ & 0.28$^{+0.14}_{-0.14}$  & 38$^{+7}_{-7}$ & \cite{Johnson12} \\
HATS-6 b & 11.2$^{+0.2}_{-0.2}$ & 101$^{+22}_{-22}$ & 0.40$^{+0.09}_{-0.09}$ & 0.20$^{+0.09}_{-0.09}$  & 29$^{+6}_{-6}$ & \cite{Hartman15} \\
TOI-3714 b & 11.3$^{+0.3}_{-0.3}$ & 222$^{+10}_{-10}$ & 0.85$^{+0.08}_{-0.08}$ & 0.1$^{+0.1}_{-0.1}$  & 47$^{+6}_{-6}$ & \cite{Canas22} \\
TOI-3235 b & 11.4$^{+0.5}_{-0.5}$ & 211$^{+8}_{-8}$ & 0.78$^{+0.11}_{-0.11}$ & 0.3$^{+0.1}_{-0.1}$  & 45$^{+6}_{-6}$ & \cite{Hobson23} \\
TOI-519 b & 11.5$^{+0.3}_{-0.3}$ & 147$^{+26}_{-28}$ & 0.56$^{+0.11}_{-0.11}$ & 0.27$^{+0.09}_{-0.09}$ & 36$^{+6}_{-6}$ & \cite{Kagetani23} \\
HATS-74 A b & 11.6$^{+0.2}_{-0.2}$ & 464$^{+44}_{-44}$ & 1.64$^{+0.19}_{-0.19}$ & 0.51$^{+0.03}_{-0.02}$  $^b$ & 73$^{+10}_{-10}$ & \cite{Jordan22} \\
TOI-5205 b & 11.6$^{+0.3}_{-0.3}$ & 343$^{+18}_{-17}$ & 1.21$^{+0.11}_{-0.11}$ & solar  & 61$^{+8}_{-8}$ & \cite{Kanodia23} \\
TOI-5293 A b & 11.9$^{+0.4}_{-0.4}$ & 170$^{+22}_{-22}$ & 0.56$^{+0.09}_{-0.09}$ & -0.03$^{+0.12}_{-0.12}$  & 40$^{+5}_{-5}$ & \cite{Canas23} \\
TOI-3757 b & 12.0$^{+0.4}_{-0.5}$ & 85$^{+9}_{-9}$ & 0.27$^{+0.05}_{-0.04}$ & 0.0$^{+0.2}_{-0.2}$  & 26$^{+4}_{-4}$ & \cite{Kanodia22} \\
TOI-1899 b & 12.9$^{+0.4}_{-0.6}$ & 210$^{+22}_{-22}$ & 0.54$^{+0.09}_{-0.10}$ & 0.28$^{+0.11}_{-0.11}$  & 45$^{+6}_{-6}$ & \cite{Lin23} \\
TOI-4201 b & $13.1^{+0.5}_{-0.5}$ & 823$^{+21}_{-20}$ & 2.0$^{+0.2}_{-0.2}$ & 0.30$^{+0.15}_{-0.15}$  & 103$^{+15}_{-15}$ & \cite{Delamer23}$^c$ \\
NGTS-1 b & 15$^{+7}_{-4}$ & 258$^{+21}_{-24}$ & 0.4$^{+0.6}_{-0.3}$ & solar  & 51$^{+7}_{-7}$ & \cite{Bayliss17}
\enddata
\vspace{0.1cm}
\textbf{Notes.}
\tablenotetext{a}{Metal masses and their errors are derived following \cite{Thorngren16} Section 5.1, which discussed a wider spread of data than the errors quoted here. We calculate the metal mass only for an order of magnitude estimate of GEMS core compositions. We performed a similar analysis following the methodology of \cite{Fortney07} and obtained consistent results.}
\tablenotetext{b}{These two metallicity measurements have small uncertainties, which only account for a part of their errors. The reported internal errors do not incorporate known systematics, which likely dominate the true uncertainty \citep{Jordan22}. }
\tablenotetext{c}{\cite{Delamer23} is quoted here and used in all relevant analyses in this work over the other two studies \citep{Gan23_4201, hartman23}. All three studies give comparable results.}
\end{deluxetable*}

TOI-5344~b is a giant planet with a radius and mass similar to Saturn. The period-radius plane (Figure~\ref{fig:planet_params} (a)) shows that TOI-5344~b and all other known transiting GEMS have relatively short periods, which can be the result of an observational bias since closer orbits have higher geometric transit probabilities. Transiting GEMS with periods shorter than half of a TESS sector ($\sim15$ days) also produce more significant detection signals in a periodogram, so TESS exoplanet searches tend to spot them more readily. We also notice the number ratio between Saturn-sized ($R_S \equiv \ \sim 8 \text{-} 10 \ \Rearth$) planets and Jupiter-sized ($R_J \equiv 10 \text{-} 15 \ \Rearth$) planets differs between the FGK-dwarf planets and M-dwarf planets. The current transiting sample shows
\begin{align}
    N_S^{\text{FGK}}/N_J^{\text{FGK}} &= 0.12\text{, and} \\
    N_S^{\text{M}}/N_J^{\text{M}} &= 0.42, 
\end{align}
where $N$ is the number of planets with FGK-dwarf or M-dwarf hosts that are Saturn-size ($S$) or Jupiter-size ($J$). The ratio difference indicates a huge lack of detection of Jupiter-size planets around M-dwarf stars (Figure~\ref{fig:planet_params} (a)). Given the small sample size of transiting GEMS, this conclusion might simply be an observational bias. However, if the conclusion holds for a comprehensive sample, several theories could explain why Saturn-size planets are more common than Jupiter-sized planets around M-dwarfs than FGK-dwarfs (Section \ref{sec:formation}).

We calculate the density of TOI-5344~b to be 0.80$^{+0.17}_{-0.15}$ g cm$^{-3}$, which is within 1$\sigma$ of Saturn's average density \citep[0.6873 $\pm$ 0.0002 g cm$^{-3}$,][]{Jacobson06}. The mass-radius plane (Figure~\ref{fig:planet_params} (b)) shows how TOI-5344~b compares to Saturn and other giant planets. A majority of the displayed giant planets (including both M-dwarf and FGK-dwarf planets) have an average density close to Saturn's density, but we only know very few Saturn-size transiting GEMS. In addition to TOI-5344~b, TOI-3629~b \citep{Canas22}, TOI-4860~b\citep{almenara23, Triaud23}, and HATS-75~b \citep{Jordan22} are three examples of similar Saturn-like planets around M dwarfs. All three planets have an early M dwarf host\footnote{HATS-75 spectral type is converted using \cite{Cifuentes20} Table 6 with the stellar mass 0.6017$^{+0.0074}_{-0.0055}$ $\Msun$ from \cite{Jordan22}. } (M1, M3.5, and M0, respectively) and super-Solar stellar metallicity\footnote{See Table \ref{tab:gems} Note b.} (0.4 $\pm$ 0.1, 0.27 $\pm$ 0.12, and 0.52$^{+0.05}_{-0.03}$, respectively). 

\subsection{Possible GEMS formation mechanisms} \label{sec:formation}

The formation of all four Saturn-like giants (TOI-5344~b, TOI-3629~b, TOI-4860~b, and HATS-75~b) around high metallicity M dwarfs suggests a possible correlation between stellar metallicity and planet density (Figure~\ref{fig:planet_params} (c)). It is crucial to note the caveats of M-dwarf metallicity derivations caused by the difference between the spectroscopic and photometric methods. While the high-resolution spectroscopic methods have generally higher precision, many stars in our sample only have SED or photometric metallicity estimation. We make no distinction between stars with measured and estimated metallicity from the NEA. Therefore, any metallicity-related discussion with the inhomogenous values from the NEA must be interpreted with caution. A positive trend is observed between planet density and stellar metallicity; the planet radii also gradually decrease with higher stellar metallicity. We perform Kendall's tau test \citep{Kendall38} between planet density and stellar metallicity for the GEMS, excluding TOI-5205~b and NGTS-1~b with only a vaguely characterized solar metallicity ([Fe/H] $\sim$ 0). We measure a correlation of $\tau = 0.5268$ and a p-value of 0.0072, which suggests a moderate correlation. 

The lowest density transiting GEMS, TOI-3757~b, orbits the second lowest stellar metallicity host star in our sample ([Fe/H] = 0.0$^{+0.2}_{-0.2}$) -- a feature discussed in detail in \cite{Kanodia22}. Conversely, the formation of GEMS is thought to benefit from the metal-rich protoplanetary disk (assumed to be similar to stellar composition), which allows the accretion of a $\sim 10 \ \Mearth$ core and comparable envelope mass \citep{Mizuno80, Pollack96} relatively fast compared to the disk dissipation speed. Current metallicity estimations of the GEMS are either Solar or super-Solar (Table \ref{tab:gems}), preferring the metal-rich disk assumption. 

We estimate the metal mass fraction of the GEMS using a metal mass-metallicity relation of \cite{Thorngren16} in Figure~\ref{fig:planet_params} (d), and TOI-5344~b has a metal mass fraction of $\sim$ 0.25, similar to Saturn \citep[0.17 $\sim$ 0.24;][]{Militzer19}. In comparison, Jupiter has a metal mass fraction of 0.057-0.103, according to a heavy-element mass of 18-33 $\Mearth$ estimated using Juno data \citep{Howard23}. The higher metal mass fraction of Saturn indicates that it failed to accrete as much gas as Jupiter did. This characteristic has led to it being referred to as a ``failed giant planet'' by many theories \citep{Helled23}, even though the exact mechanism behind this phenomenon is still debated. 

Several recent studies attempt to address the lack of gas content in Saturn and Saturn-like exoplanets. One possible explanation is that Saturns have a slowed runaway gas accretion process due to the opacity of the disk \citep{Movshovitz10}. A higher metallicity protoplanetary disk also has higher opacity, reducing the contraction heat dissipation rate of a planet. Efficient heat dissipation is necessary for rapid gas accretion to continue -- Saturns that have high metallicity disks fail to dissipate heat rapidly enough to support runaway gas accretion. \cite{Helled23} recently suggests that an intermediate stage of efficient heavy-element accretion fuels the planet with energy that hinders rapid gas accretion. In their model, a prolonged intermediate metal accretion phase extends to a few Myr at a rate of $10^{-5} \ \Mearth$ yr$^{-1}$ along with a slow gas accretion. This intermediate metal accretion provides energy to the planet and slows its accretion of gases. In addition, they suggest a runaway gas accretion at $\sim 100 \ \Mearth$, according to a turning point on the giant exoplanet Mass-Radius plane. If the disk dissipates before a planet reaches a mass $\gtrsim 100 \ \Mearth$ (still in the prolonged intermediate phase), it will fail to initiate a rapid gas accretion. This mechanism agrees with Jupiter's fuzzy core (a mixture of hydrogen, helium, and metals) of $48 - 191 \ \Mearth$ \citep{Howard23} and may explain the higher metal mass fraction of Saturn and Saturn-like exoplanets compared to Jupiters. 

These mechanisms might explain why we observe more Saturn-sized giants than Jupiter-sized giants around M-dwarf planets in comparison to FGK-dwarf planets (Section \ref{sec:params}). First, the lighter protoplanetary disk might lack enough material to form a heavier (and usually larger) planet. Also, if metal-rich is a necessity of M-dwarf hosts of giant planets like observed, the heat dissipation mechanism in high opacity would favor the formation of Saturns over Jupiters since their high metallicity disks prevent runaway gas accretion. A larger sample of GEMS is needed to confirm the lack of Jupiters around the M-dwarf stars compared to the FGK-dwarf stars.





\subsection{The planet-metallicity correlation of GEMS}
\begin{figure}
    \centering
    \includegraphics[width=\linewidth]{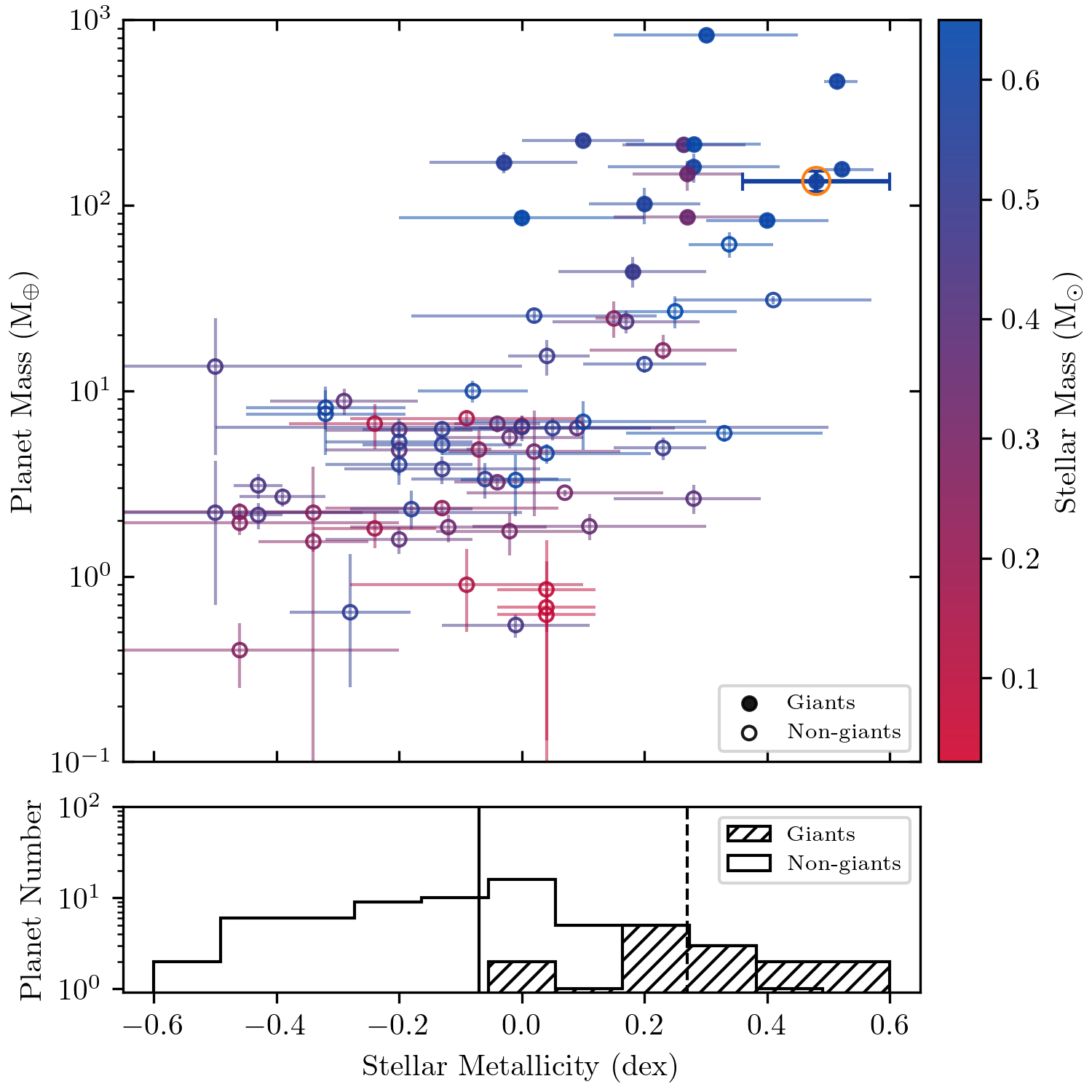}
    \caption{The planet-metallicity correlation of 15 transiting GEMS. TOI-5205~b and NGTS-1~b are excluded given their ambiguously estimated metallicities. Top: We compare transiting giants (solid circle) and transiting non-giants (empty circle) around M-dwarf stars in the Metallicity-Mass plane. The markers are color-coded with the host mass. Bottom: Histogram of the metallicity distribution. We observe the giant hosts to be more metal-rich than the non-giant hosts. The median metallicity for each population is labeled as vertical (dashed) lines. 
    }
    \label{fig:met_mass}
\end{figure}

GEMS have been theoretically expected to have only earlier type M dwarf hosts \citep{Burn21}, which agrees with current observations.
The updated transiting GEMS sample still appears to show a higher occurrence rate for earlier type M dwarfs as opposed to the non-giants ($R_p \lesssim 8 \ \Rearth$). We expect to find GEMS orbiting earlier M dwarfs because they are more massive, and have presumably more massive disks that facilitate formation via core accretion. Conversely, the non-giants have been observed with all M-type hosts, and both late-M and early-M non-giants' hosts span a wide range of metallicity (Figure~\ref{fig:met_mass}). 

The formation of giant planets around M dwarfs is considered difficult in the core accretion theory \citep{Pollack96} since the lower stellar masses usually correspond to lighter protoplanetary disks, which prevents enough solid material from being accreted before gaseous feeding zones are depleted \citep{Andrews13, Helled17}. A higher giant planet occurrence rate around metal-rich FGK-dwarf stars has been observed since the earliest discoveries of hot Jupiters \citep{Gonzalez97, Santos04, Fischer05} and later confirmed with additional samples \citep{Schlaufman14, Santos17, Narang18, Petigura18, Osborn19}.  With the caveats noted in Section~\ref{sec:params}, we revisit this hypothesis with 15 transiting GEMS (Table~\ref{tab:gems}; excluding TOI-5205~b and NGTS-1~b with imprecise metallicity estimations as ``solar''), and they all have solar or super-solar metallicity. Figure~\ref{fig:met_mass} shows transiting GEMS (solid circles) and 73 non-giant M dwarf transiting planets (empty circles) in stellar metallicity to planet mass space. Both populations are gathered from the NEA from all available surveys. While the hosts of transiting non-giants around M dwarfs are centered around Solar metallicity, transiting GEMS hosts have a median metallicity\footnote{Simply taken from the metallicity values without accounting for measurement uncertainties.} of [Fe/H] = $0.27$. 

We are tempted to statistically analyze the transiting GEMS samples against the transiting non-giants around M dwarfs in the stellar metallicity-planet mass space, but the conclusion will not be definitive given the following caveats:
\begin{enumerate}
    \item The limited sample size of transiting GEMS constrains the confidence level in potential conclusions.
    \item The comparison between the two samples (giants and non-giants) is biased since they are derived from a mixture of different populations including nearby stars from TESS and distant stars from Kepler.    
    \item The wide range of metallicity uncertainties on the stars prevents sample comparison tests such as a two-sample Kolmogorov–Smirnov (K-S) test \citep{KS_1, KS_2}, which requires homoscedasticity (all samples share similar variances).   
    \item Both samples have various metallicity measurement methods for each star. Among those with spectroscopic measurements, different instruments and reduction methods also introduce systematic errors. A fair comparison between the samples would preferably require a homogeneous reanalysis of M-dwarf stellar metallicities, which is beyond the scope of this paper. 
\end{enumerate}
We decided not to include the various statistical results we attempted on the planet-metallicity correlation due to these caveats. Two-sample K-S test \citep{KS_1, KS_2}, Kendall's tau test \citep{Kendall38}, and Anderson-Darling test \citep{AD_1,AD_2} are suitable tests to use if these caveats can be addressed. Nevertheless, Figure~\ref{fig:met_mass} suggests that the transiting GEMS hosts appear to have a different population than the transiting non-giant M-dwarf hosts. We encourage further investigation of this trend with consistent metallicity reduction and larger samples.  




\section{Conclusion}  \label{sec:summary}
We report the discovery of TOI-5344~b, a Saturn-like planet orbiting a metal-rich ([Fe/H] = $0.48 \pm 0.12$) M0 star. We discovered TOI-5344~b with TESS QLP and confirmed its planetary nature with ground-based photometry (RBO, ZTF, ASAS-SN), RVs (HPF), and speckle imaging (NESSI). 

TOI-5344~b joins a small but growing group of $\sim 20$ transiting GEMS. With a radius of $9.7^{+0.5}_{-0.5} \ \Rearth$ and a mass of $135^{+17}_{-18} \ \Mearth$, TOI-5344~b is a Saturn-like planet with an orbital period of $\sim3.79$ days. It also has a comparable metal mass fraction ($\sim0.25$) to that of Saturn, which is in a typical range (0.15 $\sim$ 0.30) shared by all 17 known transiting GEMS. 

Using the \texttt{HPF-SpecMatch} code, we show that TOI-5344 has a high metallicity of [Fe/H] = 0.48 $\pm$ 0.12, making it one of the most metal-rich transiting GEM hosts. We analyze the stellar metallicity and planet mass relation for 17 transiting GEMS and other transiting non-giants around M dwarfs and conclude they could have different host metallicity distributions. 

Ultimately, understanding GEMS requires a much larger sample size to validate existing or new theories. While GEMS are rare, the extensive grasp of TESS has yielded several new discoveries. With a gradually expanding sample size, the nature of GEMS formation will hopefully be unveiled.

\section{Acknowledgments}
Some of the data presented in this paper were obtained from MAST at STScI. Support for MAST for non-HST data is provided by the NASA Office of Space Science via grant NNX09AF08G and by other grants and contracts.
This work includes data collected by the TESS mission, which are publicly available from MAST. Funding for the TESS mission is provided by the NASA Science Mission directorate.

This research was supported in part by NASA's Exoplanet Research Program (XRP) under grant number 80NSSC23K0263.

These results are based on observations obtained with the Habitable-zone Planet Finder Spectrograph on the HET. We acknowledge support from NSF grants AST-1006676, AST-1126413, AST-1310885, AST-1310875, AST-1910954, AST-1907622, AST-1909506, ATI-2009554, ATI-2009889, ATI-2009982, AST-2108512, AST-1907622, and the NASA Astrobiology Institute (NNA09DA76A) in the pursuit of precision radial velocities in the NIR. The HPF team also acknowledges support from the Heising–Simons Foundation via grant 2017-0494. The Hobby Eberly Telescope is a joint project of the University of Texas at Austin, the Pennsylvania State University, Ludwig-Maximilians-Universität München, and Georg-August Universität Gottingen. The HET is named in honor of its principal benefactors, William P. Hobby and Robert E. Eberly. The HET collaboration acknowledges the support and resources from the Texas Advanced Computing Center. We thank the Resident astronomers and Telescope Operators at the HET for the skillful execution of our observations with HPF. We would like to acknowledge that the HET is built on Indigenous land. Moreover, we would like to acknowledge and pay our respects to the Carrizo \& Comecrudo, Coahuiltecan, Caddo, Tonkawa, Comanche, Lipan Apache, Alabama-Coushatta, Kickapoo, Tigua Pueblo, and all the American Indian and Indigenous Peoples and communities who have been or have become a part of these lands and territories in Texas, here on Turtle Island.

This work has made use of data from the European Space Agency (ESA) mission
{\it Gaia} (\url{https://www.cosmos.esa.int/gaia}), processed by the {\it Gaia}
Data Processing and Analysis Consortium (DPAC,
\url{https://www.cosmos.esa.int/web/gaia/dpac/consortium}). Funding for the DPAC
has been provided by national institutions, in particular the institutions
participating in the {\it Gaia} Multilateral Agreement.

Some of the observations in the paper made use of the NN-EXPLORE 
Exoplanet and Stellar Speckle Imager (NESSI). NESSI was funded by 
the NASA Exoplanet Exploration Program and the NASA Ames Research Center. 
NESSI was built at the Ames Research Center by Steve B. Howell, 
Nic Scott, Elliott P. Horch, and Emmett Quigley.

Based on observations obtained with the Samuel Oschin 48-inch Telescope at the Palomar Observatory as part of the Zwicky Transient Facility project. ZTF is supported by the National Science Foundation under
Grant No. AST-1440341 and a collaboration including Caltech, IPAC, the Weizmann Institute for Science, the Oskar Klein Center at Stockholm University, the University of Maryland, the University of Washington, Deutsches Elektronen-Synchrotron and Humboldt University, Los Alamos National Laboratories, the TANGO
Consortium of Taiwan, the University of Wisconsin at Milwaukee, and Lawrence Berkeley National Laboratories. Operations are conducted by COO, IPAC, and UW.

This research has made use of the SIMBAD database, operated at CDS, Strasbourg, France, and NASA's Astrophysics Data System Bibliographic Services.
This research has made use of the Exoplanet Follow-up Observation Program (ExoFOP; \dataset[DOI:10.26134/ExoFOP5]{doi.org/10.26134/ExoFOP5}) website, which is operated by the California Institute of Technology, under contract with the National Aeronautics and Space Administration under the Exoplanet Exploration Program.
This research has made use of the NASA Exoplanet Archive, which is operated by Caltech, under contract with NASA under the Exoplanet Exploration Program.

This research made use of \textsf{exoplanet} \citep{exoplanet} and its
dependencies \citep{exoplanet:agol20, exoplanet:arviz, exoplanet:astropy13,
exoplanet:astropy18, exoplanet:exoplanet, Kipping13,
exoplanet:kipping13b, exoplanet:luger18, exoplanet:pymc3, exoplanet:theano}.

CIC acknowledges support by NASA Headquarters through an appointment to the NASA Postdoctoral Program at the Goddard Space Flight Center, administered by ORAU through a contract with NASA. GS acknowledges support provided by NASA through the NASA Hubble Fellowship grant HST-HF2-51519.001-A awarded by the Space Telescope Science Institute, which is operated by the Association of Universities for Research in Astronomy, Inc., for NASA, under contract NAS5-26555. WDC acknowledges support under the NSF grant AST-2108801. RCT acknowledges support under the NSF AST-2108569.

\facilities{Gaia, HET (HPF), WIYN (NESSI), RBO, TESS, ZTF, ASAS-SN, Exoplanet Archive}

\software{
\texttt{ArviZ} \citep{exoplanet:arviz},
AstroImageJ \citep{collins_astroimagej_2017},
\texttt{astroquery} \citep{ginsburg_astroquery_2019},
\texttt{astropy} \citep{exoplanet:astropy13, exoplanet:astropy18},
\texttt{barycorrpy} \citep{Kanodia18},
\texttt{celerite2} \citep{foreman-mackey_fast_2017, foreman-mackey_scalable_2018}
\texttt{exoplanet} \citep{exoplanet, exoplanet:exoplanet},
\texttt{HxRGproc} \citep{HxRGproc},
\texttt{ipython} \citep{perez_ipython_2007},
\texttt{lightkurve} \citep{lightkurve_collaboration_lightkurve_2018},
\texttt{matplotlib} \citep{hunter_matplotlib_2007},
\texttt{numpy} \citep{oliphant_numpy_2006},
\texttt{pandas} \citep{mckinney_data_2010},
\texttt{pyastrotools} \citep{kanodia_shubham_2023}
\texttt{PyMC3} \citep{exoplanet:pymc3},
\texttt{scipy} \citep{oliphant_python_2007, virtanen_scipy_2020},
\texttt{SERVAL} \citep{Zechmeister18},
\texttt{starry} \citep{exoplanet:luger18, exoplanet:agol20},
\texttt{tglc} \citep{TGLC},
\texttt{Theano} \citep{exoplanet:theano},}


\bibliography{toi-5344}{}
\bibliographystyle{aasjournal}
\end{document}